\documentclass[prc,aps,showpacs]{revtex4}
\usepackage[russian]{babel}
\usepackage{amssymb}
\usepackage{epsfig}     
\usepackage{wrapfig}
\usepackage{graphics}
\usepackage{amssymb}      

\def\g{\gamma}
\def\a{\Gamma}

\def\pp{\pi^+}

\def\po{\pi^0}

\def\go{\g d\to\po\g pn}
\def\beq{\begin{eqnarray}}
\def\eeq{\end{eqnarray}}

\begin{document}
                
\title{Supernarrow Dibaryons }

\author{L.V.~Fil'kov\footnote[1]{filkov@sci.lebedev.ru} 
\vspace*{0.3cm}}

\affiliation{P.N. Lebedev Physical Institute RAS, Leninskiy 
Prospect 53, Moscow, 119991, Russia}

\begin{abstract}
An analysis of the experimental search for supernarrow dibaryons (SNDs)
have been performed. The sum rules for SND masses have been constructed.
The calculated values of the SND masses are in good agreement with the
existing experimental values. It has been shown that the SND decay leads
to the formation of $N^*$ with small masses. Experimental observations of 
$N^*$ is an additional confirmation of the possibility of the SND existence.  

\end{abstract}


\maketitle

\section{Introduction}

Supernarrow dibaryons (SNDs) are 6-quark states, a decay of which into
two nucleons is forbidden by the Pauli exclusion principle 
\cite{fil1,fil2,alek1}.
Such states satisfy the following condition:
\begin{equation}
(-1)^{T+S}P=+1
\end{equation}
where $T$ is the isospin, $S$ is the internal spin, and $P$ is the
dibaryon parity. In the $NN$ channel, these six-quark states would
correspond to the following forbidden states:
even singlets and odd triplets with the isotopic spin $T=0$ as well as
odd singlets and even triplets with \mbox{$T=1$}.
These six-quark states with the masses \mbox{$M < 2m_{N}+m_{\pi}$}
($m_N (m_{\pi})$ is the nucleon (pion) mass) can mainly decay
by emitting a photon. This is a new class of metastable six-quark states
with the decay widths \mbox{$< 1$keV}. 

The experimental discovery of the SNDs would have important consequences
for particle and nuclear physics and astrophysics. This would lead to
a deeper understanding of the evolution of compact stars and the new
possibility of quark-gluon plasma observation. 
In nuclear physics there would be a new concept: SND-nuclei.

In the framework of the MIT bag model, Mulders et al. \cite{muld} calculated
masses of different dibaryons, in particular, masses of $NN$-decoupled 
dibaryons. They predicted dibaryons $D(T=0;J^P =0^{-},1^{-},2^{-};M=2.11$ GeV)  
and $D(1;1^{-};2.2$ GeV) corresponding to the forbidden states $^{13}P_J$
and $^{31}P_1$ in the $NN$ channel.
However, the dibaryon masses obtained exceed the pion production threshold.
Therefore, these dibaryons preferentially decay into the $\pi NN$ channel
and their decay widths are larger than 1 MeV.

Using the chiral soliton model, Kopeliovich \cite{kop} predicted that
the masses of \mbox{$D(1,1^+)$} and $D(0,2^+)$ SNDs could
exceeded the two nucleon mass by 60 and 90 MeV, respectively.
These values are lower than the pion production threshold.

In the framework of the canonically quantized biskyrmion model,
Krupnovnickas {\em et al.} \cite{riska} obtained an indication on
possibility
of the existence of one dibaryon with $J=T=0$ and two dibaryons with $J=T=1$
with masses smaller than $2m_N+m_{\pi}$.

In the present paper we analyze the experimental search for SNDs, construct
and analyze the mass formula for SNDs, and suggest a possible interpretation
of exotic narrow baryons with low masses.

\section{Supernarrow dibaryons}
\label{sec:2}

We will consider the following SNDs:
$D(T=1,J^P=1^+,S=1)$ and $D(1,1^-,0)$.

It is worth noting, that the  state $(T=1, J^P=1^-)$ corresponds
to the states $^{31}P_1$ and $^{33}P_1$ in the NN channel.
The former is forbidden and
the latter is allowed for a two-nucleon state. In our work we will study
the dibaryon $D(1,1^-,0)$, a decay of which into two nucleons is forbidden
by the Pauli principle (i.e. $^{31}P_1$ state).

SNDs can be formed in processes of interaction with the deuteron
only if the  nucleons in the deuteron overlap sufficiently, such that a
6-quark state with deuteron quantum numbers can be formed. In this case, an
interaction of a photon or a meson with this state can change its
quantum numbers so that a metastable state can form.
Therefore, the probability of the production of such dibaryons is
proportional to the probability $\eta$ of the 6-quark state existing
in the deuteron.

The magnitude of $\eta$ can be estimated from the
deuteron form factor at large $Q^2$ (see for example \cite{bur,grach}). However,
the values obtained depend strongly on the model of the form factor of
the 6-quark state over a broad region of $Q^2$. Another way to estimate
this parameter is to use the discrepancy between the theoretical and
experimental values of the deuteron magnetic moment \cite{kim,bhad,kon2}.                    
This method is free from the restrictions quoted above and gives
$\eta\le 0.03$ \cite{kon2}.

\begin{wrapfigure}{h}{0.3\textwidth}
\epsfxsize=0.3\textwidth      
\epsfysize=2cm
\epsffile{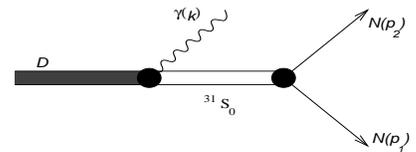}   
\caption{The diagram of the SND decay into $\g NN$}
\label{dec}
\end{wrapfigure}
Since the energy of nucleons, produced in the decay of the SNDs under
study with $M<2m_N +m_{\pi}$, is small, it is expected that the main
contribution to a two nucleon system should come from the
$^{31}S_0$ (virtual singlet) state (Fig. \ref{dec}).

\begin{table}
\label{table1}
\centering
\caption{ Decay widths of the dibaryons $D(1,1^+,1)$ and $D(1,1^-,0)$
at various dibaryon masses $M$.} 
\begin{tabular}{|c|c|c|c|c|c|c|}\hline
$M$(GeV)     &1.904  &1.926 &1.942 &1.965 &1.985  &2.006  \\ \hline
$\a_t(1,1^-)$&0.0514 &0.327 & 0.771&1.909 & 3.495 &5.073  \\
(eV)         &       &      &      &      &       &       \\ \hline
$\a_t(1,1^+)$&0.206  &1.307 &3.083 &7.635 &13.98  &23.49  \\
(eV)         &       &      &      &      &       &        \\ \hline
\end{tabular}
\end{table}
The results of calculations of the
decay widths of the dibaryons into $\g NN$ on the
basis of such assumptions at $\eta=0.01$ are listed in Table I.

As a result of the SND decay through $^{31}S_0$ in
the intermediate state, the probability distribution of such a decay
over the emitted photon energy $\omega$ should be characterized by a
narrow peak at the photon energy close to the maximum value                                   
$\omega_m=(M^2-4m^2_N)/2M$ (Fig.~2).                              
Note that the interval of the photon energy from $\omega_m$ to
$\omega_m-1$ MeV contains about 75\% of the contribution to the width
of the decay $D(1,1^{\pm})\to \gamma NN$. This leads to a very small
relative energy of the nucleons from the SND decay              
and these nucleons are emitted into a narrow angle cone with respect
to the direction of the SND motion. 
\begin{figure}[h]
\epsfxsize=6cm      
\epsfysize=6cm
\centering{
\epsffile{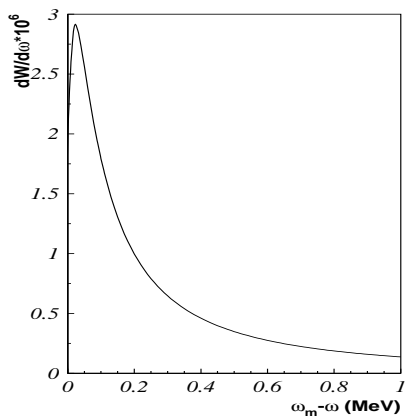}}     
\caption{The distribution of the decay probability of the isovector
SND over $\omega_m-\omega$ at $M=1904$ MeV.}
\label{width}
\end{figure} 

\begin{figure}[ht]
\epsfxsize=6cm     
\epsfysize=6cm
\centering{
\epsffile{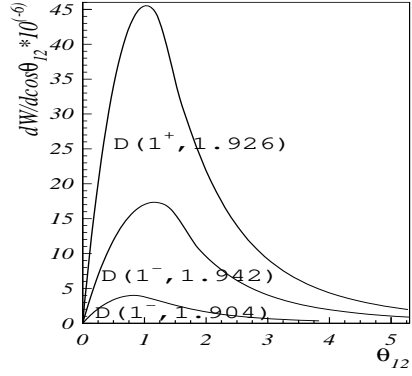}}        
\caption{The distribution of the decay probability of SNDs over 
$\theta_{1,2}$}
\label{d-angl}
\end{figure}     
Moreover, the distribution of the SND decay probability over the angle
between the final nucleons should be characterized by a narrow angular
cone. 

The dependence of SND decay probability on the angle
between the final nucleons $\theta_{12}$ is shown in Fig.~3. 
This figure  demonstrates that the nucleons from 
the decay of SNDs should be mainly emitted in a very 
narrow angular cone. 
These dependencies should be taken into account when looking for SNDs. 

\section{Previous works}

For the first time, SNDs have been searched for in the reactions
$pd\to p+pX_1$ and $pd\to p+dX_2$
\cite{konob,izv,yad,prc,conf1,conf2,epj,hadr}.
The experiment was
carried out at the Proton Linear Accelerator of INR with 305 MeV
proton beam using the two-arm mass
spectrometer TAMS. 

Several software cuts have been applied to the mass spectra in these works.
In particular, the authors limited themselves by the consideration of 
intervals of the proton energy and angles from the decay of the $pX_1$ 
states, and very narrow angular cone between final nucleons, which
were determined by the kinematics of the SND decay into $\gamma NN$ channel.
Such cuts are very important as it provides a possibility to suppress the
contribution from the background reactions and random coincidences
essentially.

In the works \cite{conf1,conf2,epj,hadr}, CD$_2$ and $^{12}$C were
used as targets.
The scattered proton was detected in the left arm
of the spectrometer TAMS at the angle $\theta_L=70^{\circ}$. The second
charged particle (either $p$ or $d$) was detected in
the right arm by three telescopes located at $\theta_R=34^{\circ}$,
$36^{\circ}$, and $38^{\circ}$.

\begin{figure}[ht]                    
\epsfxsize=7cm     
\epsfysize=9cm
\centering{
\epsffile{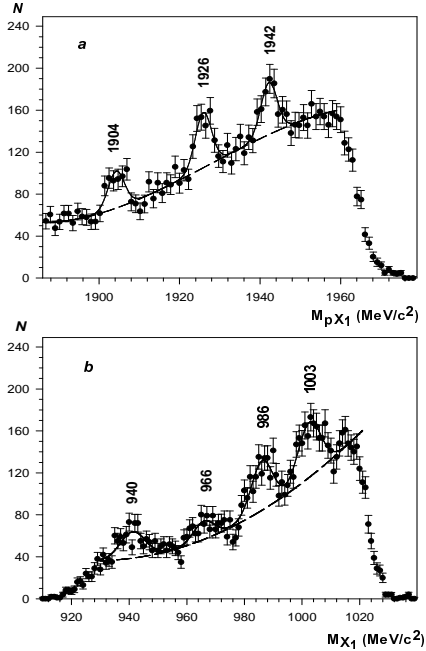}}              
\caption{The missing mass $M_{pX_1}$ (a) and
$M_{X_1}$ (b) spectra of the reaction $pd\to p+pX_1$.}
\label{inr}
\end{figure}    
As a result,
three narrow peaks in missing mass spectra have been observed 
(Fig.\ref{inr}) at $M_{pX_1}=1904\pm 2$, $1926\pm 2$, and $1942\pm 2$ MeV  
with widths equal to the experimental resolution ($\sim 5$ MeV) and 
with numbers of standard deviations (SD) of 6.0, 7.0, and 6.3, respectively.
It should be noted that the dibaryon peaks at $M=1904$ and 1926 MeV had
been observed earlier by same authors in ref. \cite{prc,konob,izv,yad}
at somewhat different kinematical conditions. 

On the other hand, no noticeable
signal of the dibaryons has been observed in the missing mass $M_{dX_2}$
spectra of the reaction $pd\to p+dX_2$.
The analysis of the angular distributions of the protons from the decay of
$pX_1$ states and the suppression observed of the SND decay into $\gamma d$
showed that the peaks found can be explained as a manifestation of the
isovector SNDs, the decay of which into two nucleons is forbidden
by the Pauli exclusion principle.

An additional information about the nature of the observed states has been
obtained by studying the missing mass $M_{X_1}$ spectra of the
reaction $pd\to p+pX_1$.
If the state found is a dibaryon decaying mainly into two nucleons then
$X_1$ is the neutron and the mass $M_{X_1}$ is equal to the neutron mass
$m_n$. If the value of $M_{X_1}$, obtained from the experiment, differs
essentially from $m_n$, then $X_1=\gamma+n$ and we have the additional
indication that the observed dibaryon is SND.

The simulation of the missing mass $M_{X_1}$ spectra of the reaction
$pd\to ppX_1$ has been performed \cite{conf1,conf2,epj,hadr} assuming that
the SND decays as $D\to\g+\, ^{31}S_0\to \g pn$ through two nucleon singlet
state $^{31}S_0$ \cite{fil2,prc,epj}. As a result, three narrow peaks at
$M_{X_1}=965$, 987, and 1003 MeV have been predicted. These peaks
correspond to the decay of the isovector SNDs with masses 1904, 1926, and
1942 MeV, respectively.

In the experimental missing mass $M_{X_1}$ spectrum besides the peak
at the neutron mass due to the process $pd\to p+pn$,
a resonance-like behavior of the spectrum has been observed at $966\pm 2$,
$986\pm 2$, and $1003\pm 2$ MeV \cite{conf1,conf2,epj,hadr}.
These values of $M_{X_1}$ coincide with
the ones obtained by the simulation and essentially differ from
the value of the neutron mass (939.6 MeV). Hence, for all states under
study, we have $X_1=\gamma+n$ in support of the statement that the
dibaryons found are SNDs.

On the other hand, the peak at $M_{X_1}=1003\pm 2$ MeV corresponds to
the peak found at SPES3 (Saturne) ref. \cite{tat2} and was attributed to an 
exotic baryon state $N^*$ below the $\pi N$ threshold. In that work the 
authors investigated the reaction $pp\to\pi^+pX$ and have found
altogether three such states with masses 1004, 1044, and 1094 MeV 
with SD$\gtrsim 10$. In additional, states with masses close to 966 and
986 MeV were also extracted at SPES3, but  from a small  number of
data \cite{tat3}.

Therefore, if the exotic baryons with anomalously small masses really
exist, the observed peaks at 966, 986, and 1003 MeV might be a manifestation
of such states. The existence of such exotic states, if
proved to be true, will fundamentally change our understanding of the
quark structure of hadrons \cite{bald,walch,azim}.

However, in experiments on a single nucleon, no any significant structure 
was observed \cite{lvov,jiang,kohl}. Therefore,
the question about a nature of the peaks observed in \cite{epj,tat2} 
remains open at present.                      

In ref. \cite{khr} dibaryons with exotic quantum numbers were searched for
in the process $pp\to pp\gamma\gamma$. The experiment was performed with
a proton beam from the JINR Phasotron at an energy of about 216 MeV. The
energy spectrum of the photons emitted at $90^{\circ}$ was measured.
As a result, two peaks have been observed in this spectrum. This behavior
of the photon energy spectrum was interpreted as a signature of the exotic
dibaryon resonance $d_1^*$.  
with a mass of about $1956\pm 2 stat\pm 6 syst$ MeV and possible isospin 
$T=2$ or $T=1$.

On the other hand, an analysis \cite{cal} of the Uppsala proton-proton
bremsstrahlung data looking for the presence of a dibaryon in the mass 
range from 1900 to 1960 MeV gave only the upper limits of 10 and 3 nb for
the dibaryon production cross section at proton beam energies of 200 and
310 MeV, respectively.

However, it is asserted in ref. \cite{khr2} that a more detailed analysis 
of the data of ref \cite{cal} gives a confirmation of the existence of 
$d_1^*(1956)$.
 
\begin{figure}[ht]   
\epsfxsize=6cm       
\epsfysize=4cm
\centering{
\epsffile{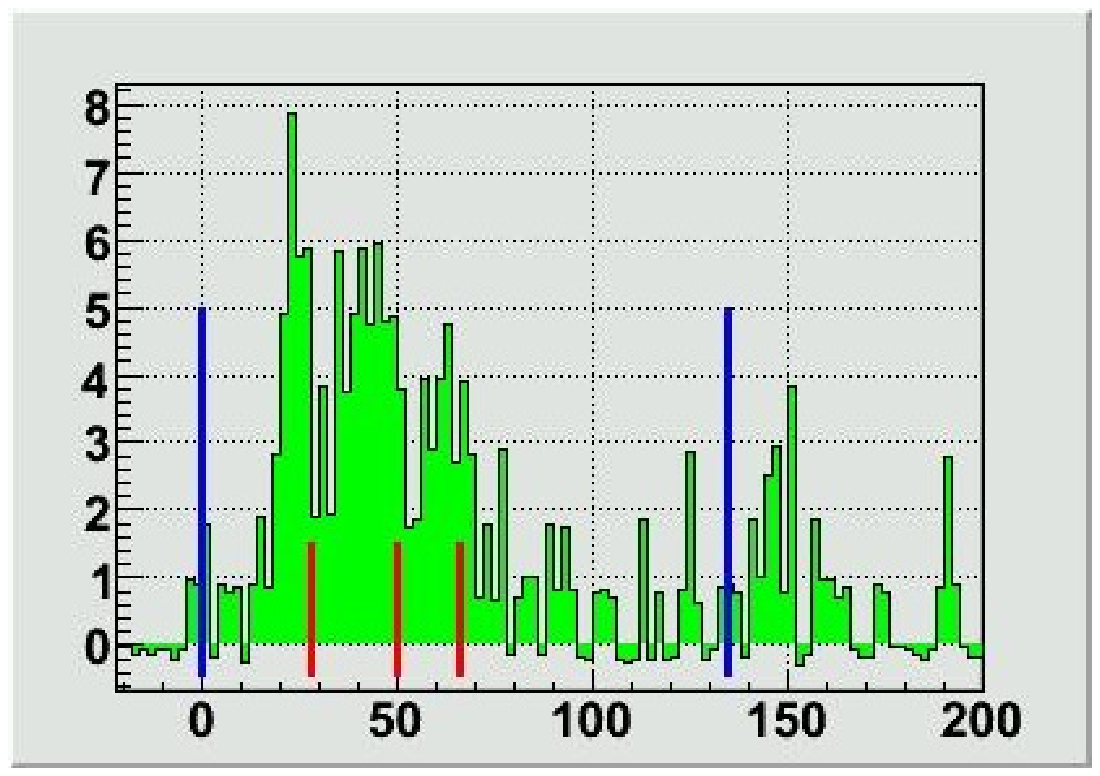}}    
\caption{The missing mass $(MM(\g,\pi^0)-m_d)$ (MeV)		 
spectra of the reaction $\go$ }
\label{mami}
\end{figure}      
The reactions $pd\to ppX$ and $pd\to pdX$ have been studied also 
in the Research Center for Nu\-clear Physics at the pro\-ton energy 295 MeV 
over a mass range of 1898 to 1953 MeV \cite{tamii}. They did not observe 
any narrow structure in the missing mass spectra of $pX$, $dX$, and $X$.

So, these results are at variance both with the SND observation in INR and
with the results of Tatischeff {\em et al.} \cite{tat2,tat3} on the search for
exotic baryons. However, exotic baryons were observed in Ref. \cite{tat2}
with a sufficiently high accuracy, which leads to a doubt about correctness
of the result of RCNP \cite{tamii}. 

It is worth noting that the reaction $pd\to NX$ was investigated in
other works, too. However, in contrast to
the ref. \cite{konob,izv,yad,prc,conf1,conf2,epj,hadr},
the authors of these works did not study
either the correlation between the parameters of the scattered proton and
the second detected particle
or the emission of the photon from the dibaryon decay.
Therefore, in these works the relative contributions of the dibaryons under
consideration were small, which hampered their observation.

On the other hand, the preliminary analysis of the missing mass distributions 
of the available data on the process $\go$, obtained at MAMI \cite{kash}, 
demonstrates three peaks (Fig. \ref{mami}), which good enough correspond to 
the values of the SND masses found in INR \cite{prc,epj} (red lines). 
Unfortunately, the statistic is very poor in this case.

\begin{figure}[ht]      
\epsfxsize=6cm     
\epsfysize=5cm
\centering{
\epsffile{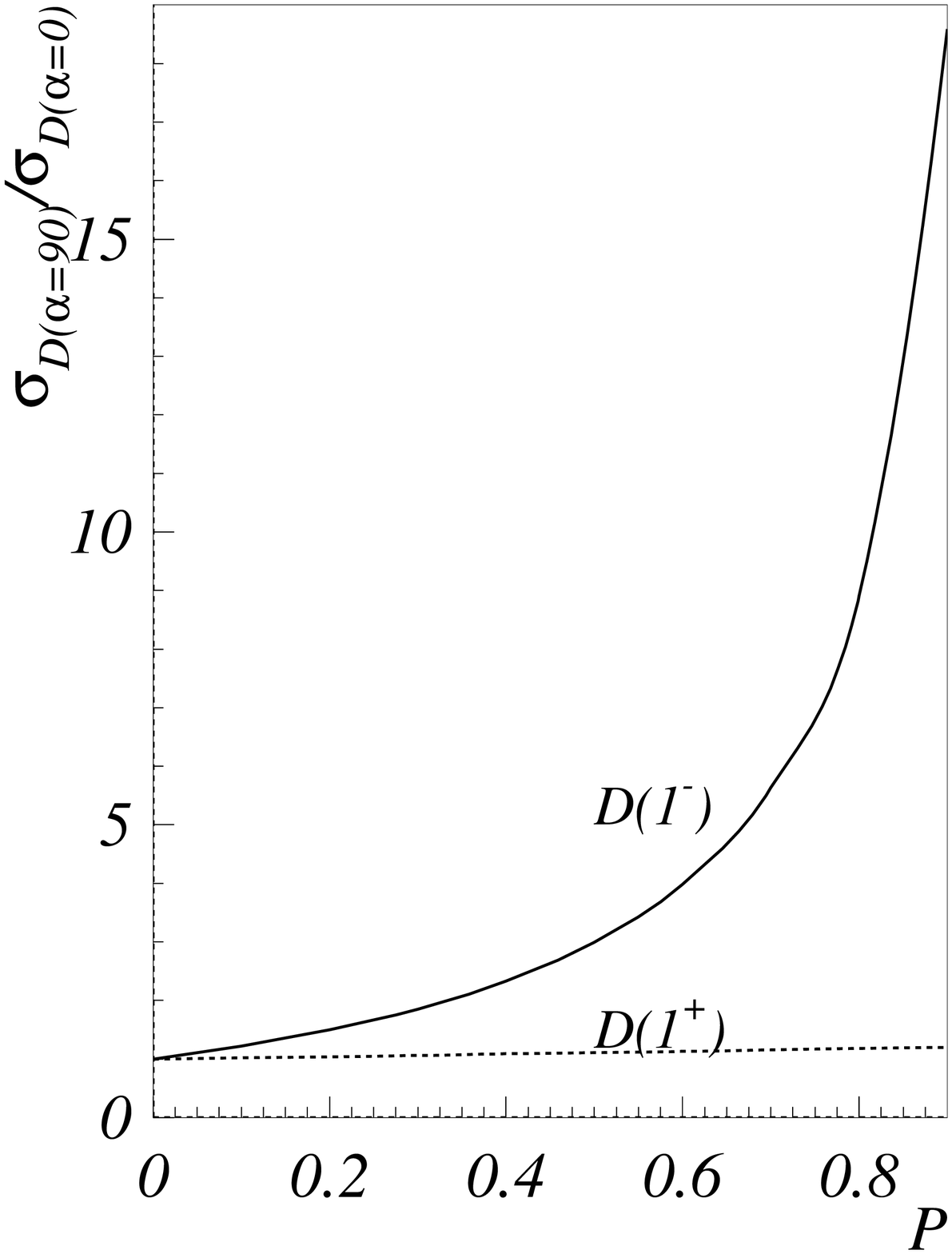}}       
\caption{The re\-la\-tio of $\sigma (\alpha=90^o)/\sigma (\alpha=0^o)$ 
as the function of $P$.}
\label{polar}
\end{figure}      
In order to argue more convincingly that the states found are
really SNDs, an additional experimental investigation of the dibaryon
production is needed.

In ref. \cite{alek1} a search for SNDs in the processes of pions
photoproduction by the linearly polarized photon was proposed. 
The cross section of this process can be written as the following:
\beq
 \frac{d\sigma}{d\Omega}=A+\frac{q^2}{2}\sin^2\theta_{\pi} \, B
(1-P \,\cos 2\alpha) 
\eeq
where $\alpha$ is the angle of the photon polarization relative to
the reaction plane.

The result of calculations of $\sigma (\alpha=90^o)/\sigma (\alpha=0^o)$ 
for the process $\overrightarrow{\g} d\to \pi^+ D$ at $M=1904$ MeV is shown
on Fig. \ref{polar} as the function of the polarization degree $P$.

So, we have a large variation between the differential cross sections of 
pions propagating parallel to the initial photon polarization and pions 
propagating perpendicular to the polarization. It is expected that the
cross section for vector SNDs should be substantially larger for mesons
emitted parallel to the photon polarization than for mesons emitted
perpendicularly. Thus, in the first case, the sensitivity to the contribution
of $D(1,1^-)$ significantly increased.

\begin{figure}[ht]       
\epsfxsize=6cm      
\epsfysize=10cm      
\centering{
\epsffile{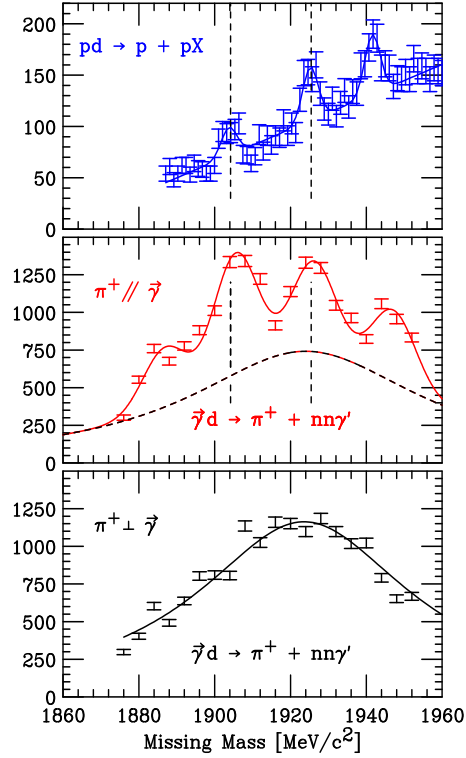}}      
\caption{The missing mass spectra of the reactions $pd\to p+ pX$,
and $\vec{\gamma} d\to \pi^+nn\gamma$ when $\pi^+//\vec{\gamma}$
and $\pi^+\perp\vec{\gamma}$.}
\label{norum}
\end{figure}                                                                            

The calculations showed that $D(1,1^-)$ gives the main contribution
to the amplitude $B$, and its contribution to $A$ is less than 1\%.
The contributions of $D(1,1^+)$ in $A$ and $B$ are almost equal. 
If we consider the SND masses up to 1960 MeV, then it is
expected that when a pion is emitted parallel to the photon
polarization, 3 peaks should be observed and only 1, when the pion 
is emitted perpendiculary.

Such an experiment has been performed at LEGS \cite{legs,norum}. They
analyzed the reaction $d(\overrightarrow{\g},\pp n\g')n$ in the photon                               
energy range 210 -- 340 MeV. The linear polarization of initial photon
was approximately 99\%. The results of this experiment and their comparison
with the data obtained at INR \cite{prc,epj} are shown in Fig. 7. 

As a result they have observed three peaks in missing mass spectrum
when the $\pp$ was emitted parallel to the polarization of the incident
photon $\overrightarrow{\g}$. 

The mass values found are very close 
to the values obtained in INR \cite{prc,epj} 

The peak at $M=1926$ MeV in the bottom panel corresponds to the expected
value. However, this peak is too wide, possibly due to insufficient
accuracy in the determination of experimental data in the region of this
resonance. 

The results obtained in this experiment,
support the possibility of the SND existence. However, these data were
limited by the resolution of the pion detection. So, they did not
produce a conclusive proof of the SND existence. 

It should be noted that SNDs could be produced in the processes under 
consideration, if a pion is only emitted from the 6-quark state of the
deuteron. Therefore the vertexes of $d\to \pi D$ can be written as
\beq 
&&\a_{d\to\pi D(1,1^{-},0)}=\frac{g_1}{M} \sqrt{\eta}
\Phi_{\mu \nu} G^{\mu \nu} ,\\
&&\a_{d\to\pi D(1,1^+,1)}=\frac{g_2}{M}\sqrt{\eta}\varepsilon_{\mu\nu
\lambda\sigma}\Phi^{\mu\nu}G^{\lambda\sigma},\\ \nonumber
\eeq
where $\Phi_{\mu \nu}=r_{\mu}w_{\nu}-w_{\mu}r_{\nu}$,
$G_{\mu \nu}=p_{1\mu}v_{\nu}-v_{\mu}p_{1\nu}$, $w$ and $v$ are 4-vectors
of the dibaryon and deuteron polarizations, respectively; and 
$r$ and $p_1$ are the dibaryon and deuteron 4-momenta.

The constants $g^2_1/4\pi$, $g^2_2/4\pi$, and $\eta$ are unknown.
However, the products of these coupling constants and $\eta$ can be
estimated from the results of work \cite{legs} where the SNDs,
were searched for in the process $\overrightarrow{\g} d \to \pp D\to\g'\pp nn$.

 As a result, we have
\begin{equation}
\eta\frac{g_1^2}{4\pi}=1.4\times 10^{-4}, \qquad
\eta\frac{g_2^2}{4\pi}=3\times 10^{-4}.
\end{equation}

\section{Mass formula for the SNDs}

Using the complete Green function of the dibaryons
$$
\Delta(p^2)=\frac{F(p)}{p^2-m^2-\delta_D(p^2)},
$$
we determine the SND mass as
\begin{equation}\label{mdb}
M^2=m^2+Re\,\delta_D(M^2),
\end{equation}
where $\delta_D(M^2)$ is the self energy of the SND under study and
$m$ is the mass of the dibaryon in the intermediate state.

The self energy of the lightest SND will be determined in one loop
approximation through the interaction of the pion with the  6-quark
state of the deuteron.
The self energy of the next SND will be obtain through the interaction of
the pion with the lightest SND and so on.

We calculate the SND self energy with the help of
the dispersion relations with two subtractions at $M^2=m^2$.
Then taking into account (\ref{mdb}), we obtain the mass formula for the
SNDs \cite{mass}
\begin{eqnarray}
\label{dsd}
M^2&=&m^2+Re\,\delta_D(m^2)+\left.(M^2-m^2)\frac{
d\,Re\,\delta_D(M^2)}{d\,M^2}\right|_{M^2=m^2}+ \nonumber \\
&&\frac{(M^2-m^2)^2}{\pi}P\!\!\!\int\limits_{(m+\mu)^2}^{\infty}
\frac{Im\,\delta_D(x)\,dx}{(x-M^2)(x-m^2)^2}.
\end{eqnarray}
Since the subtraction is carried out on the mass shell of the
dibaryon in the intermediate state, the subtraction function
$Re\,\delta_D(m^2)$ is equal to zero. Assuming that this dibaryon is
in the ground state, we have
$\left.d\,Re\,\delta_D(M^2)/d\,M^2\right|_{M^2=m^2}=0$.

Finally, the mass formula for SND can be represented as
\begin{equation}
\label{dsd1}
\frac{(M^2-m^2)}{\pi}P\int\limits_{(m+\mu)^2}^{\infty}
\frac{Im\,\delta_D(x)\,dx}{(x-M^2)(x-m^2)^2}=1   \nonumber
\end{equation}

Two subtractions in the dispersion relations provide a very good
convergence of the integrand in (\ref{dsd1}). Therefore we restrict
ourselves to consideration of one loop approximation only.

We assume that the SND under study and the dibaryon in the intermediate
state have opposite parities. 
Then the vertex $D^{\prime}(1^{\mp})\to\pi+D(1^{\pm})$  
can be written as
\begin{equation}
\Gamma^{(-)}=\frac{\bar{g}_1}{M}G_{\mu\nu}\Phi^{\mu\nu}. \qquad
\end{equation}

As a result of  calculations we have got the following expression for
the imaginary part of $\delta_D(x)$: 
\begin{equation}\label{oppos}
Im\,\delta_D(x)=\left(\frac{\bar{g}^2_1}{4\pi}\right)
\frac{q[(x+m^2-\mu^2)^2+2m^2x]}{x^{\frac32}},
\end{equation}
where $q$ is the pion momentum equal to
$q=[(x-(m+\mu)^2)(x-(m-\mu)^2]^{1/2}/2x^{1/2}$.

The coupling constant $\bar{g}^2_1/4\pi$ in the vertex for transition of
the 6-quark state of the deuteron $(D(0,1^+))$ plus the pion to the
SND $D(1,1^-)$ has been fixed by requiring a reproduction of the mass
$M=1904$ MeV. It results in
\begin{equation}\label{const1}
\frac{\bar{g}^2_1}{4\pi}=26.5888 .
\end{equation}

Calculations within the framework of the present model yielded very close 
values of the SND masses found in channels with $\pi^0$ and 
$\pi^{\pm}$ mesons . Therefore, we take them equal one to another.

In order to calculate the mass of the next SND $D(1,1^+)$, we take in the
intermediate state the SND $D(1,1^-)$ with $m=1904$ MeV and the pion.
To calculate the mass of the next $D(1,1^-)$, we consider SND 
$D(1,1^+)$ with $m=1925$ MeV and pion in the intermediate state and so on. 

The coupling constant $\bar{g}_2$ in the vertex
$D(1,1^{\pm})\to D(1,1^{\mp})+\pi$ differs from $\bar{g}_1$.
Due to the isotopic invariance, we have
$\bar{g}^2_2/4\pi=3/4 (\bar{g}^2_1/4\pi)$. 

The results of the calculations of the masses
and $J^P$ of the SNDs are listed in table \ref{dibm}.
\begin{table}
\centering
\caption{The masses and $J^P$ of the SNDs.}
\label{dibm}
\begin{tabular}{|c|c|c|c|c|}\hline
     &     &  model  &  experiment & experimental \\
$No$& $J^P$&$M$ (MeV)& $M$ (MeV)   &  works       \\ \hline
 1   &$1^-$& 1904    &$1904\pm 2$  & \cite{epj}   \\ \hline
 2   &$1^+$& 1925    &$1926\pm 2$  & \cite{epj}   \\ \hline
 3   &$1^-$& 1945    &$1942\pm 2$  & \cite{epj}   \\ \hline
 4   &$1^+$& 1965    &$1956\pm 6$  & \cite{khr}   \\ \hline
 5   &$1^-$& 1985    & $1982$ &predicted \cite{epj,tat2}\\ \hline
 6   &$1^+$& 2006    &            &               \\ \hline
\end{tabular}
\end{table}

As can be seen from the table, the values of the SND masses obtained are in
good agreement with available experimental data. The existence of the SND 
with the mass $M=1982$ MeV was predicted in \cite{epj,tat2}
as a consequence of the observation of the peak in the missing mass 
spectrum of the reaction $pp\to\pi^+pX$ \cite{tat} at $M_X=1044$ MeV.

An analysis of the probability of SND decay shows that due to the
smallness of the angle $\theta_{12}$ and difference in the energies
of the final nucleons, resonance-like states with  masses
$(p_1+k)^2\simeq (p_2+k)^2=M_{N^*}^2$ appear.
The table \ref{nstar} presents the values of $M_{N^*}$ for 
various SND masses. 
\begin{table}
\centering
\caption{The masses of the SNDs and $N^*$.}
\label{nstar}
\begin{tabular}{|c|c|c|c|c|c|c|}\hline
$J^P$&$1^-$ &$1^+$ & $1^-$ & $1^+$ & $1^-$ & $1^+$ \\ \hline
 M(D)&$1904\pm 1$&$1925\pm 1$&$1945\pm 1$&$1965\pm 1$&$1985\pm 1$
 &$2006\pm 1$\\ \hline
$M(N^*_{SR})$&$965\pm 2$&$986\pm 2$&$1005\pm 2$&$1025\pm 1$&$1044\pm 1$ 
 &$1063\pm 1$\\ \hline
$M(N^*_{exp})$&$966\pm 2$&$986\pm 2$&$1003\pm 2$& &$1044\pm 2$& \\
      &INR\cite{epj}&INR\cite{epj}&INR\cite{epj}& &SPES3\cite{tat2}& \\
              &  &SPES3\cite{tat3} &SPES3\cite{tat2}  & &           & \\
\hline
\end{tabular}
\end{table}

As the SNDs observed in \cite{prc,epj} decay into $NN^*$ with the small
relative momentum between $N$ and $N^*$, the SND parity has to be
determined by the parity of the $N^*$. As seen from tables II and III,
the SND parities found here agree with the results obtained for
the $N^*$ states.

Thus, the formation of $N^*$ is uniquely determined by the existence of
SNDs. $N^*$ is not a new particle. This is a kinematic effect 
associated with the SND decay form. Consequently, the experimental
observation of $N^*$s can be an additional indication of the possibility 
of the existence of SNDs.
                                                  
\section{Summary}
                                                  
The main properties of SNDs have been considered. A number of experiments
\cite{prc,epj,khr,kash,legs}, where evidence was obtained for the 
existence of SNDs, has been analyzed. 
                                             
Negative results obtained in RCNP \cite{tamii} are at variance both with
the observation of SNDs in INR \cite{prc,epj} and with the result of
B. Tatischeff {\em et al.} \cite{tat2} on search for exotic baryons. On the
other hand, the latter was observed in \cite{tat2} with sufficiently 
high accuracy, which leads to doubt about correctness of the result of
RCNP \cite{tamii}.

The sum rules for the SND masses have been constructed. The values  of
the SND masses obtained by means of the sum rules agree very well with
the experimental data \cite{epj,tat2}.

The decay of SNDs leads to the formation of resonance-like states $N^*$.
The predicted values of the masses of $N^*$ are in good agreement with
available experimental data.

The experimental observation of $N^*$s is an additional conformation 
of the possibility of the SND existence.
 
The author thanks V.L. Kashevarov and B. Norum for helpful discussion.

\end{document}